\documentclass[preprint,preprintnumbers,amsmath,amssymb,floatfix]{revtex4}

\usepackage{graphicx}
\def\be{\begin{equation}}
\def\ee{\end{equation}}
\def\lf{\left (}
\def\rt{\right )}

\begin{document} 

\preprint{IUCAA-09/2008}
\preprint{ITFA-2008-11}

\title{Not One Bit of de Sitter Information}

\author{Maulik Parikh$^{1}$ and Jan Pieter van der Schaar$^{2}$}

\affiliation{$^{1}$Inter-University Centre for Astronomy and Astrophysics, 
Post Bag 4, Pune 411007, India \\
$^{2}$ Institute for Theoretical Physics, University of Amsterdam, The Netherlands
\bigskip
\bigskip
\bigskip}

\begin{abstract}
We formulate the information paradox in de Sitter space in terms of the no-cloning principle of quantum mechanics. We show that energy conservation  
puts an upper bound on the maximum entropy available to any de Sitter observer. Combined with a general result on the average information in a quantum subsystem, this guarantees that an observer in de Sitter space cannot obtain even a single bit of information from the de Sitter horizon, thereby preventing any observable violations of the quantum no-cloning principle, in support of observer complementarity. 
\end{abstract}

\maketitle
\thispagestyle{empty}

The black hole information paradox remains one of the most important challenges in theoretical physics \cite{hawkinginfopuzzle}. Although much progress has been made over the years, a definitive answer on how black hole information is conserved and possibly retrieved is still lacking. 
Moreover, since the presence of an event horizon seems to be the crucial ingredient in formulating the information paradox, the problem 
could extend beyond black holes, in particular to de Sitter space. 

If the principle of equivalence holds, an in-falling observer crossing a black hole horizon would notice nothing remarkable there. But if black hole evaporation is a unitary process the Hawking radiation left after the hole has disappeared must encode a faithful copy of any quantum system that the in-falling observer was carrying. Hence, at earlier times, two copies of the state seem to be needed, one inside and one outside the black hole. But this seems to fall afoul of the no-cloning principle of quantum mechanics. This principle (which complicates the task of making backups in quantum computing) says that no unitary operator can make duplicate copies of a general quantum state; copying violates linearity. A proposed resolution to this dilemma is the principle of black hole complementarity which in effect permits the cloning to take place so long as no observer can witness it \cite{stu, susskindcomp}. This profound and highly nontrivial proposal looks easy to falsify at first sight but a number of careful thought experiments indicate that complementarity cannot be ruled out, at least for black holes \cite{larussusskind,thooft,compl}.

In pure de Sitter space, it is less clear that an information puzzle can be formulated \cite{theswedes}. Nevertheless, because de Sitter radiation can be derived in much the same way that black hole radiation can \cite{Gibhaw,newcoords,medved}, one expects that de Sitter radiation too contains information about things that have fallen through the horizon. But now we again run the risk of observable violations of the no-cloning principle. The purpose of this note is to show that these do not occur: energy conservation protects observer complementarity and no observer observes any duplicate information. In fact, we will derive the stronger result that an observer in de Sitter space can not retrieve even a single bit of information. 

It helps to first recall a few facts about information.
The information contained in a system is the deficit between the maximal coarse-grained entropy that it could have and the entropy that it actually has:
\be
I = S_{\rm maximal} - S_{\rm actual} \; . \label{info}
\ee
Intuitively, the more disordered a system is, the closer it is to thermal equilibrium and maximal entropy, and therefore the less information it contains. A calculation of the information contained in a subsystem of a larger system was done in a remarkable paper by Don Page \cite{donpage}. Page imagined that the total system was in a pure quantum state within the total Hilbert space, ${\cal H}$.  Let ${\cal H}$ be divided into a tensor product of two Hilbert spaces of dimension $m$ and $n$, corresponding respectively to the Hilbert spaces of the subsystem and of the rest of the system. Tracing over the states of the rest of the system yields a density matrix, $\rho$, for the subsystem. Page calculated the entanglement entropy, $-{\rm tr} \rho \ln \rho$, for the subsystem, and averaged it over all possible pure states for the total system (where the averaging was with respect to a uniform weighting for pure states given by the unitarily invariant (Haar) measure). The result \cite{donpage,pageproof}, for $m \leq n$, is
\be
\left < S \right > = \sum_{k = n+1}^{k = mn} {1 \over k} - {m - 1 \over 2n} \; . \label{page}
\ee
Since $S_{\rm maximal} = \ln m$,  the average information in the subsystem is, by (\ref{info}) and (\ref{page}),
\be
\left < I_{m,n} \right > = \ln m - \left < S \right > \approx {m \over 2n} \; , \label{infosub}
\ee
where the approximation assumes that $m, n \gg 1$. When $m = n$, the subsystem and the rest of the system each contain half the entropy. Even so, (\ref{infosub}) says that the information contained individually within each of them is typically just half a unit; the bulk of the information is encoded in the correlations between the two parts of the system. Since the information in even a single bit is $\ln 2$, we see that effectively no information is contained in the subsystem until it contains more states than half of the total number of states in the system. That is, for a typical pure state in the total system a subsystem carries no information unless it has at least half the total entropy of the system.
\begin{figure}[hbtp]
 \centering
  \includegraphics[angle=0,width=80mm]{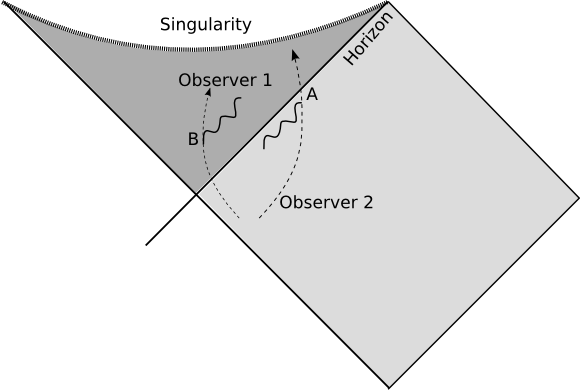}
 \centering
 \caption{Scenario for potential violation of black hole complementarity.}
 \label{fig:bhc}
\end{figure}

These general considerations have had profound consequences for the black hole information puzzle. Consider the situation shown in the Penrose diagram of Figure \ref{fig:bhc}. Observer 1 falls into a black hole carrying with him a quantum system, a spin degree of freedom say. Observer 2 stays outside until, at point A, she has received the Hawking radiation carrying the information about the spin state. She then bravely jumps into the black hole to try to see the same bit twice. Meanwhile observer 1 must send a message to observer 2 about the spin no later than point B in order for the message to get to observer 2 before she hits the singularity. Page's calculation is essential here because it says that half the entropy of the hole must be radiated out before any information about the spin state can emerge. In $D$ spacetime dimensions, an information retention time of order
\be
t_{\rm info} \sim \lf G_D^2 M^{D-1} \rt^ {1 \over D-3} 
\ee
must have elapsed before observer 2 can jump into the hole at point A. The delayed in-fall of observer 2 leaves a limited amount of time for observer 1 to send a signal; it can then be shown that on the basis of the uncertainty principle the energy required to send the message would back-react severely on the geometry, thereby invalidating the entire semi-classical set-up \cite{larussusskind,susskindbook}. 

\begin{figure}[hbtp]
 \centering
\includegraphics[angle=0,width=60mm]{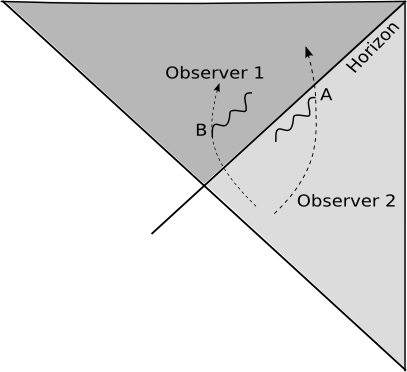}
 \centering
 \caption{Scenario for potential violation of de Sitter complementarity.}
 \label{fig:dsc}
\end{figure}

For de Sitter space, the Penrose diagram looks quite similar (see Figure \ref{fig:dsc}), but there are a few differences. The first obvious difference is the absence of a singularity. 
More crucially, the analogue of the information retention time is obscure for de Sitter space. Without a delayed in-fall, the observer who records the Hawking radiation could potentially jump through the horizon and still have enough time to receive the message with the duplicate bit.

Now, in a strong version of holography, the finite entropy of de Sitter space actually enumerates the logarithm of the number of states in the finite Hilbert 
space ${\cal H}$ of de Sitter quantum gravity \cite{bankscc,ellipticdS,dsholography,toystory,qdeformeddS,albertodavid,giddingsmarolf}. In $D$ spacetime dimensions, the Bekenstein entropy is $\Omega_{D-2} L^{D-2}/4G_D$ where $G_D$ is Newton's constant, $\Omega_{D-2}$ is the volume of a unit $D$-sphere, and $L$ is the radius of curvature of de Sitter space. We regard the Hilbert space ${\cal H}$ of the entire system to have dimension equal to the exponential of this number. Now we would like to estimate how much information is available to an observer inside a de Sitter horizon. As (\ref{infosub}) indicates, the information content depends on the entropy of the subsystem available to the observer. In particular, the observer is immersed in a bath of de Sitter radiation. What is its entropy? 
A gas of blackbody radiation with ${\cal N}$ polarizations, or ${\cal N}$ massless degrees of freedom in thermal equilibrium, in de Sitter space has entropy
\be
S = {4 {\cal N} D \over (4 \pi L)^{D-1}} {\Gamma(D-1) \zeta(D) \over \lf \Gamma \lf { D -1 \over 2} \rt \rt^2} \left [ \int_0^{L-\epsilon} {r^{D-2} dr \over \lf 1 - (r/L)^2 \rt^{D/2}} \right ] \; .
\ee
Here we have used the usual blackbody entropy density at the local temperature and integrated it over the de Sitter horizon volume. This expression diverges if the ultraviolet cut-off $\epsilon$ is removed because of the $D$ factors of $\sqrt{1-(r/L)^2}$ in the denominator of the integrand ($D-1$ of which come from the blueshifted temperature and one coming from the proper volume). In fact, even with an ultraviolet cut-off, the entropy in the blackbody radiation can be arbitrarily large if the number of massless degrees of freedom is unrestricted. Naively, it appears that the observer could have access to 
an almost unlimited amount of information.

However, this estimate is no good; we have neglected gravity. Indeed, it is now understood that it is precisely gravitational backreaction, or energy conservation, that allows the quanta to tunnel across the de Sitter horizon \cite{newcoords,tunnel,secret,energy}. But when including gravity, the entropy can be 
determined by considering the resulting de Sitter-Schwarzschild black hole, instead of the thermal de Sitter radiation. Consider then the line element 
for Schwarzschild-de Sitter space:
\be
ds^2 = - \lf 1 - {2G_D M \over r^{D-3}} - {r^2 \over L^2} \rt dt^2  +   \lf 1 - {2G_D M \over r^{D-3}} - {r^2 \over L^2} \rt^{\!-1} dr^2 + r^2 d \Omega^2_{D-2} \; .
\ee
For $M < M_{\rm max}$, this has two horizons, a cosmological horizon at radius $r_+$ and, within it, a black hole horizon of radius $r_-$. At $M = M_{\rm max}$, the two horizons coincide at the radius $r_H$; the solution is sometimes called the Nariai black hole. The maximum amount of entropy that can be stored in de Sitter space is clearly the entropy of a Nariai black hole. To find its mass $M_{\rm max}$ we use the fact that the temperature of a de Sitter black hole is
\be
T_{\rm hole} = {1 \over 2 \pi} \lf {(D-3) G_D M \over r_-^{D-2}} - {r_- \over L^2} \rt \; .
\ee
As with extremal black holes in asymptotically flat space, the Nariai black hole has vanishing temperature, which is in fact a general consequence of coincident horizons. We thus obtain $G_D M_{\rm max} = (D-3) r_H^{D-1} /L^2$. Using $g_{00}=0$ at the horizon we find
\be
G_D M_{\rm max} = {L^{D-3} \over D-1} \lf {D-3 \over D-1} \rt^{\! (D-3)/2}
\ee
and obtain
\be
r_H = \lf { D-3 \over D -1} \rt^{\! 1/2} L \; .
\ee
Thus we find
\be
\label{finalresult}
{S_{\rm subsystem} \over S_{dS}} \leq \lf { D - 3\over D - 1} \rt^{\! (D-2)/2} < {1 \over e} \; ,
\ee
which is bounded above by $e^{-1}$, in the limit of infinite $D$. We see that, for all values of $D > 3$, the entropy in the Nariai black hole is less than half the total entropy of de Sitter space. Thus even if all the Hawking radiation were most efficiently captured and stored in a massive black hole, the maximum entropy of the subsystem accessible to the static observer would still be insufficient to encode even a single bit of information. 

The spontaneous creation of these massive black holes in de Sitter space violates the second law of thermodynamics because the total horizon area decreases from its maximum value (given by the entropy of empty de Sitter space). Nevertheless, such thermodynamic fluctuations are to be expected, and occur in far less time than a random Poincare recurrence. This is relatively straightforward to see. Poincare recurrences occur on a time-scale of the order of the de Sitter entropy (in Planck units), $\tau_P \sim e^{S_{dS}}$. Instead, on general grounds the probability for Hawking radiating a black hole equals $\Gamma \sim e^{\Delta S}$, where $\Delta S = S_{f}-S_{i}$ is the change in the entropy of the cosmological horizon before and after emission of the black hole \cite{newcoords}. Since the cosmological and black hole horizons coincide for the Nariai black hole the final de Sitter entropy equals the black hole entropy, which was calculated above in (\ref{finalresult}). Using that result we end up with the following estimate for the average time-scale $\tau_N$ (in Planck units and for $D>3$) to produce a Nariai black hole
\be
\label{bhprod}
\tau_N \sim \frac{1}{\Gamma} \sim \exp{\left[ \left(1-{S_{\rm Nariai} \over S_{dS}} \right) S_{dS}\right]}  \lesssim \tau_P^{\; \frac{2}{3}} \ll \tau_P\; .
\ee
Although this probability is tiny, the time-scale is clearly much smaller than the recurrence time. But what we have shown above is that even these extreme fluctuations do not provide any information for an observer to set up a violation of the no-cloning principle. To be precise, one assumption that goes into this is that the original pure state is a typical one. Since we know that the pure Bunch-Davies (or Euclidean) vacuum actually corresponds to an entangled thermal state to any free-falling observer, this seems to be a valid assumption. 

We have only considered the scenario of an observer hoping to cross the horizon after recording the Hawking radiation. But it is easy to see that observers who stay on one side or other of the horizon also do not meet with any contradictions. Indeed, all hot horizons can be approximated by Rindler horizons and arguments showing that Rindler observers can be reconciled with complementarity were already made in \cite{larussusskind}. 

To conclude, using energy conservation and a basic result in quantum information theory, we have shown that it is impossible for any observer in de Sitter space to measure even a single bit of information. Consequently, de Sitter complementarity appears to be safe. 

\vspace{0.5cm}
\noindent
{\bf Acknowledgments:}
We would like to thank Daniel Kabat, Erik Verlinde and Jan de Boer for useful discussions. M.~P. thanks the Department of Physics at Columbia University, where part of this work was done. J.~P.~v.d.~S. would like to thank the Inter-University Centre for Astronomy and Astrophysics in Pune, India for its hospitality during part of this work. The research of J.~P.~v.d.~S. is financially supported by Foundation of Fundamental Research on Matter (FOM) grant 06PR2510.


\begin{thebibliography}{99} 

\bibitem{hawkinginfopuzzle}
S.~W.~Hawking, ``Breakdown of Predictability in Gravitational Collapse,"
Phys. Rev. D {\bf 14} (1976) 2460.

\bibitem{stu}
  L. Susskind, L. Thorlacius, and J. Uglum, ``The Stretched Horizon
  and Black Hole Complementarity,'' Phys. Rev. {\bf D48} (1993) 3743;
  {\tt hep-th/9306069}.
  
\bibitem{susskindcomp}
L.~Susskind, ``String theory and the principles of black hole complementarity,"
Phys. Rev. Lett. {\bf 71} (1993) 2367; 
{\tt hep-th/9307168}.

\bibitem{larussusskind}
L.~Susskind and L. Thorlacius, ``Gedankenexperiments involving black holes,"
Phys. Rev. D. {\bf 49} (1994) 966; 
{\tt hep-th/9308100}.

\bibitem{thooft}
  C.R. Stephens, G. 't Hooft, and B. F. Whiting. ``Black Hole Evaporation
  Without Information Loss," Class. Quant. Grav. {\bf 11} (1994) 621;
  {\tt gr-qc/9310006}.

\bibitem{compl} 
  Y. Kiem, E. Verlinde, and H. Verlinde, ``Black Hole Horizons and
  Complementarity," Phys. Rev. D {\bf 52} (1995) 7053; {\tt hep-th/9502074}.

\bibitem{theswedes}
U.~H.~Danielsson, D.~Domert, and M.~E.~Olsson, ``Miracles and complementarity in de Sitter space," Phys. Rev. D {\bf 68} (2003) 083508; {\tt hep-th/0210198}.

\bibitem{Gibhaw}
  G. Gibbons and S. Hawking, ``Cosmological Event Horizons, Thermodynamics,
  and Particle Creation,'' Phys. Rev. {\bf D15} (1977) 2738.

\bibitem{newcoords}
M.~Parikh, ``New Coordinates for de Sitter Space and de Sitter
Radiation,'' Phys. Lett. B {\bf 546} (2002)
189; {\tt hep-th/0204107}.

\bibitem{medved}
A. J. M. Medved, ``Radiation via Tunneling from a de Sitter Cosmological Horizon,'' Phys. Rev. D {\bf 66} (2002) 124009; {\tt hep-th/0207247}.

\bibitem{donpage}
D.~N.~Page, ``Average Entropy of a Subsystem," Phys. Rev. Lett. {\bf 71} (1993) 1291; {\tt gr-qc/9305007}.

\bibitem{pageproof}
S.~K.~Foong and S.~Kanno, ``Proof of Page's conjecture on the average entropy of a subsystem," Phys. Rev. Lett. {\bf 74} (1994) 1148.

\bibitem{susskindbook}
L.~Susskind and J.~Lindesay, {\it Black Holes, Information, and the String Theory Revolution}
(World Scientific, 2005).

\bibitem{bankscc}
T.~Banks, ``Cosmological breaking of supersymmetry? or Little lambda goes back to the future 2," {\tt hep-th/0007146}.

\bibitem{ellipticdS}
M.~Parikh, I.~Savonije, and E.~Verlinde, ``Elliptic de Sitter Space: $dS/\mathbb{Z}_2$,'' 
Phys. Rev. D {\bf 67} (2003) 064005;
{\tt hep-th/0209120}.

\bibitem{dsholography}
M.~K.~Parikh and E.~P.~Verlinde, ``De Sitter holography with a number of states," JHEP {\bf 0501} (2005) 054; {\tt hep-th/0410227}.

\bibitem{toystory}
M.~K.~Parikh and E.~P.~Verlinde, ``De Sitter space with finitely many states: A toy story," {\tt hep-th/0403140}.

\bibitem{albertodavid}
A.~Guijosa and D.~A.~Lowe, ``A New twist on dS / CFT," 
Phys. Rev. D {\bf 69} (2004) 106008;
{\tt hep-th/0312282}.

\bibitem{qdeformeddS}
D.~A.~Lowe, ``q-deformed de Sitter / conformal field theory correspondence,"
Phys. Rev. D {\bf 70} (2004) 104002;
{\tt hep-th/0407188}.

\bibitem{giddingsmarolf}
S.~B.~Giddings and D.~Marolf, ``A global picture of quantum de Sitter space," Phys. Rev. D {\bf 76} (2007) 064023; {\tt hep-th/0705.1178}.



\bibitem{energy}
M.~K.~Parikh, ``Energy conservation and Hawking radiation," {\tt hep-th/0402166}.
 
\bibitem{secret}
M.~K.~Parikh, ``A Secret tunnel through the horizon," 
Int. J. Mod. Phys. D {\bf 13} (2004) 2351; Gen. Rel. Grav. {\bf 36} (2004) 2419;
{\tt hep-th/0405160}.
 
\bibitem{tunnel}
M. K. Parikh and F. Wilczek, ``Hawking Radiation as Tunneling,'' Phys. Rev. Lett. {\bf 85} (2000) 5042; {\tt hep-th/9907001}.

\end{thebibliography}
\end{document}